# XANES study of rare-earth valency in $L$Ru$_4$P$_{12}$ ($L=$ Ce and Pr)

C. H. Lee and H. Oyanagi
*Electrotechnical Laboratory, 1-1-4 Umezono, Tsukuba, Ibaraki 305-8568, Japan*

C. Sekine and I. Shirotani
*Muroran Institute of Technology, 27-1, Mizumoto, Muroran 050, Japan*

M. Ishii
*JASRI, Kamigori, Ako-gun, Hyogo 678-12, Japan*

Valency of Ce and Pr in $L$Ru$_4$P$_{12}$ ($L=$ Ce and Pr) was studied by $L_{2,3}$-edge x-ray absorption near-edge structure (XANES) spectroscopy. The Ce-$L_3$ XANES spectrum suggests that Ce is mainly trivalent, but the $4f$ state strongly hybridizes with ligand orbitals. The band gap of CeRu$_4$P$_{12}$ seems to be formed by strong hybridization of $4f$ electrons. Pr-$L_2$ XANES spectra indicate that Pr exists in trivalent state over a wide range in temperature, 20 K$\leq T\leq$300 K. We find that the metal-insulator (MI) transition at $T_{\text{MI}}=60$ K in PrRu$_4$P$_{12}$ does not originate from Pr valence fluctuation. [S0163-1829(99)01244-8]

## I. INTRODUCTION

Ce and Pr compounds show a wide variety of phenomena such as heavy fermion, valence fluctuation, and Kondo effect due to the instability of $4f$ electrons. It is well known that $4f$ electrons of Ce and Pr are at the borderline between localization and itinerancy.

Recently, skutterudite compound PrRu$_4$P$_{12}$ has attracted great attention due to the occurrence of MI transition at $T_{\text{MI}}=60$ K.[1] PrRu$_4$P$_{12}$ is metallic at room temperature and becomes semiconductive below $T_{\text{MI}}$. The Rietveld analysis of the powder x-ray diffraction on PrRu$_4$P$_{12}$ shows no structural phase transition from cubic $Im\bar{3}$ below room temperature down to $T=10$ K.[1] Magnetic susceptibility does not show distinct anomaly at $T=T_{\text{MI}}$.[1] The MI transition is driven by neither change in crystal symmetry nor magnetic long-range ordering. Mechanism of the MI transition in PrRu$_4$P$_{12}$ is still controversial.

Other skutterudite compounds $L$Ru$_4$P$_{12}$ ($L=$ La, Ce, Nd, Sm, Gd, and Tb) are metallic except for $L=$ Ce and Sm, which show semiconducting transport property and MI transition at $T_{\text{MI}}=16$ K, respectively.[2–4] Crystal structure at room temperature is the same in all $L$Ru$_4$P$_{12}$ compounds. Susceptibility measurements show that SmRu$_4$P$_{12}$ is antiferromagnetic (AF) with a Néel temperature of $T_N=16$ K. In SmRu$_4$P$_{12}$, the semiconducting property below $T_{\text{MI}}$ seems to originate from AF long-range ordering. Basically, $L$Ru$_4$P$_{12}$ compounds are metallic, and the semiconducting phase in CeRu$_4$P$_{12}$ and in PrRu$_4$P$_{12}$ below $T_{\text{MI}}$ without magnetic ordering is exotic. It is quite natural to consider that the $4f$ electron instability plays an important role for the semiconducting property.

For a band calculation in skutterudite compounds, there are two models assuming either localized or itinerant $f$ electrons. According to the band calculation with completely localized La$^{3+}$ cations in LaFe$_4$P$_{12}$, the top of $t_{2g}$ bands is separated from the bottom of $e_g$ bands by a band gap.[5] When rare earth is trivalent, the highest occupied $t_{2g}$ band is half-filled so that metallic behavior is expected. On the other hand, with tetravalent rare earth, $t_{2g}$ bands are completely filled, and transport property should be semiconducting. Another band calculation in semiconducting CeFe$_4$P$_{12}$ taking itinerancy of $f$ electrons into account also shows a band gap around the Fermi level.[6] In such a case, the band gap is the result of strong hybridization of Ce $4f$ with Fe $3d$ and P $3p$ orbitals. In the latter calculation, valency of the Ce atom is predicted to be nearly trivalent contrary to the former calculation, which anticipates tetravalent Ce for semiconducting CeRu$_4$P$_{12}$. Thus, determination of Ce valency is important to judge which is the proper model for CeRu$_4$P$_{12}$.

Being based on the latter itinerant model, PrRu$_4$P$_{12}$ is always metallic since the total number of electrons is odd and it is difficult to explain the MI transition. On the other hand, in the former localized model, the MI transition in PrRu$_4$P$_{12}$ can occur due to fluctuation of Pr valency between trivalent and tetravalent states. Studying the temperature dependence of Pr valency of PrRu$_4$P$_{12}$ is, thus, very important.

X-ray absorption near-edge structure (XANES) is now a well-established method of determining the valency of rare earth. A number of previous studies has shown that tetravalent Ce and Pr ions have double-peaked white lines in their $L_{2,3}$-XANES spectra, while trivalent Ce and Pr ions have only a single-peaked white line.[7–10]

In this paper, we report the results of XANES measurements in $L$Ru$_4$P$_{12}$ ($L=$ Ce and Pr) compounds. The aim of the present work is to elucidate the mechanism of MI transition in PrRu$_4$P$_{12}$ and the origin of semiconducting behavior in CeRu$_4$P$_{12}$ by studying Pr and Ce valency.

## II. EXPERIMENTAL DETAIL

$L$Ru$_4$P$_{12}$ ($L=$ Ce and Pr) samples were synthesized by a high-pressure cell under high temperature using a wedge-type cubic-anvil high-pressure apparatus. The obtained powders were characterized to be a single phase using powder x-ray diffraction. Resistivity measurements reveal MI transition at $T_{\text{MI}}=60$ K for PrRu$_4$P$_{12}$. CeRu$_4$P$_{12}$ behaves as a semiconductor with an activation energy $\Delta\omega=0.074$ eV. Details of synthesis and characterization are given in Refs. 1, 2, 4, and 11.

XANES measurements were performed in a fluorescence mode at BL13B station of Photon Factory on the Pr $L_2$-edge and at BL10XU of SPring-8 on the Ce $L_3$ edge. At the Photon Factory, we have used a 27-pole wiggler magnet[12] and a Si (111) double crystal monochromator, which is directly water-cooled.[13] In order to avoid the degradation of energy resolution due to distortion of Si monochromator as a result of heat load, incident x-ray intensity was suppressed to 2/3 of the full power by limiting the magnetic field of the 27-pole wiggler magnet to $B_0 = 1.0$ T where $B_0 = 1.5$ T provided the maximum beam intensity. The fluorescence yield was measured by an array of 19-element pure Ge solid-state detectors.[14] The output of a single-channel analyzer for each detector was recorded and normalized by the incident-beam intensity measured by an ionization chamber filled with dry nitrogen gas. Energy resolution was about 2.0 eV at the Pr-$L_2$ threshold (at around 6.44 keV). Powder samples of $PrRu_4P_{12}$ were pressed into pellets, and they were mounted on an aluminum holder of a closed-cycle He refrigerator. The stability of temperature was $\pm 0.1$ K.

At SPring-8, we used an x-ray undulator radiation from an in-vacuum undulator (U32V) (Ref. 15) and a rotated-inclined double crystal Si (111) monochromator.[16] Since intensity distribution of undulator radiation in the energy range is narrow, we tuned the undulator gap twice for each XANES measurement so that the incident x-ray intensity does not fall off below 75% of maximum intensity. Energy resolution was about 2.0 eV at the Ce-$L_3$ threshold (at around 5.73 keV). Details of experimental setup are given elsewhere.[17] Powder samples of $CeRu_4P_{12}$ were sandwiched by Kapton tapes, and they were mounted on an aluminum holder.

The measured XANES spectra were normalized by the standard procedure. A linear background estimated from the pre-edge was subtracted from the XANES data, and they were normalized to unit intensity at about 25 eV above the absorption edge. For energy calibration, the spectrum of Cu $K$ edge was recorded. The repeatability of the incident x-ray energy was checked to be better than 0.2 eV by measuring the Cu $K$-edge spectra for several times.

### III. RESULTS

Figure 1(a) shows the Ce $L_3$-edge XANES spectra of $CeTiO_3$ and $CeO_2$ measured as standard materials of trivalent and tetravalent Ce compounds, respectively. As is shown, single peak for trivalent and double peak for tetravalent Ce have been observed clearly. Peak $A$ in trivalent Ce corresponds to excitation from $2p_{3/2}$ to empty $5d$ states with final configuration $\underline{2p}4f^15d^*$ ($\underline{2p}$ and $5d^*$ denote a hole in $2p$ state and an excited electron in $5d$ state, respectively). In tetravalent Ce, according to many-body calculations with Anderson impurity model,[18–20] the double peaks arise from many-body final states. Final-state configuration for peaks $B$ and $C$ is $\underline{2p}4f^1\underline{L}5d^*$ and $\underline{2p}4f^05d^*$, respectively ($\underline{L}$ denotes a hole in ligand orbitals).

Ce $L_3$-edge XANES spectrum of $CeRu_4P_{12}$ is shown in Fig. 1(b). The spectrum does not have simple single nor double peak structure as $CeTiO_3$ and $CeO_2$, but has a well-defined peak $D$ with a weak satellite peak $E$. We defined peak position at the maximum intensity and edge energy at 1/3 of the maximum signal. Such determined peak position and edge energy of all measured samples are summarized in Table I. Position of peak $D$ agrees with the main peak of $CeTiO_3$. Furthermore, the energy position of absorption threshold of $CeRu_4P_{12}$ is also almost the same with that of $CeTiO_3$. Agreement in energy positions and the fact that the satellite peak is weak suggest that the valency of Ce in $CeRu_4P_{12}$ is mainly trivalent.

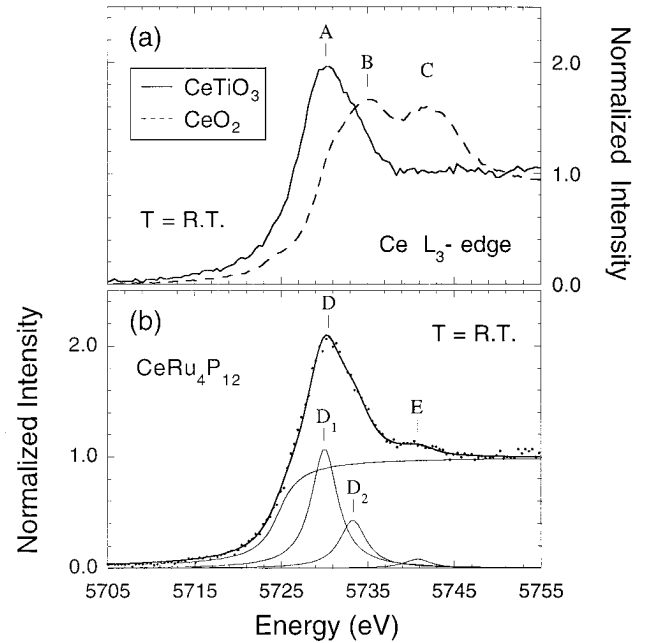

FIG. 1. Ce $L_3$-edge XANES spectra of (a) $CeTiO_3$, $CeO_2$, and (b) $CeRu_4P_{12}$ at room temperature. The thick solid line in (b) is a result of fitting using three Lorentzian functions and an arctan function convoluted with a Gaussian function. Lorentzian and arctan components are depicted by thin solid lines.

Figure 2 shows Pr $L_2$-edge XANES spectra of $PrRu_4P_{12}$ taken at $T = 20$, 60, and 300 K. As is shown, well-defined peak $F$ is observed without a satellite peak. Peak $F$ corresponds to excitation with final-state configuration $\underline{2p}4f^25d^*$. The spectra at three temperatures agree quite well with each other. It seems that position and linewidth of peak $F$ do not have a temperature dependence. The well-defined single peak structure independent of temperature suggests trivalent state of Pr over a wide range of temperature (20 K $\leq T \leq$ 300 K).

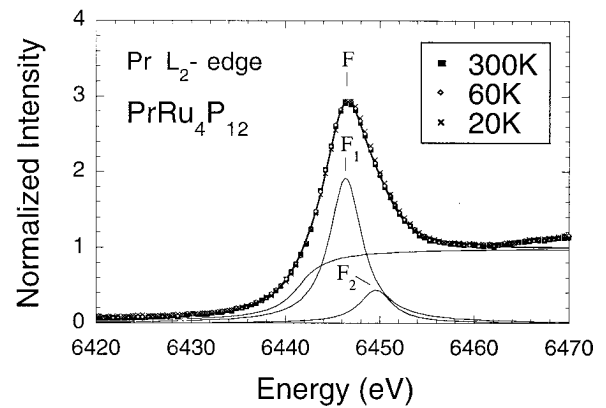

FIG. 2. Pr $L_2$-edge XANES spectra of $PrRu_4P_{12}$ at $T = 300$, 60, and 20 K. The thick solid line is a result of fitting using two Lorentzian functions and an arctan function convoluted with a Gaussian function. Lorentzian and arctan components are depicted by thin solid lines.

TABLE I. Parameters of the Ce and Pr $L_{2,3}$-edge spectra. The first inflection point in spectrum was determined to be the edge energy. $\varepsilon_1$, $\Gamma$, and $\delta$ are the results of fitting using Eq. (1).

| | Edge energy (eV) | Peak position (eV) | $\varepsilon_1$ (eV) | $\Gamma$ (eV) | $\delta$ (eV) |
|---|---|---|---|---|---|
| Ce $L_3$ edge | | | | | |
| CeTiO$_3$ | 5725.0(5) | A 5730.5(5) | | | |
| CeO$_2$ | 5728.0(5) | B 5735.0(5) | | | |
| | | C 5742.5(5) | | | |
| CeRu$_4$P$_{12}$ | 5725.0(5) | D 5730.5(5) | $D_1$ 5730.0(2) | 3.7(5) | 5724.7(3) |
| | | | $D_2$ 5733.3(2) | 3.7(5) | |
| | | E 5740.8(4) | E 5740.8(4) | | |
| Pr $L_2$ edge | | | | | |
| PrRu$_4$P$_{12}$ | 6441.7(5) | F 6446.5(5) | $F_1$ 6446.4(2) | 4.2(5) | 6441.5(3) |
| | | | $F_2$ 6449.6(2) | 4.2(5) | |

To study configuration of peaks in both CeRu$_4$P$_{12}$ and PrRu$_4$P$_{12}$ more carefully, we fitted the spectra using the following function convoluted with a Gaussian resolution function of which full width at half maximum (FWHM) is 2.0 eV,

$$F(\omega) = \sum_i \frac{a_i}{(\omega - \varepsilon_i)^2 + (\Gamma/2)^2} + \left[0.5 + \frac{1}{\pi}\arctan\left(\frac{\omega - \delta}{\Gamma/2}\right)\right], \quad (1)$$

where $\omega$, $\varepsilon_i$, $\Gamma$, $\delta$, and $a_i$ are photon energy, peak position for excitation into $5d$ bound states, core-hole lifetime, excitation energy into continuum states, and constant value, respectively. The Lorentzian and the arctan function describe the transitions to bound states and to continuum, respectively. As shown in Fig. 1(b) and in Fig. 2, the fitted lines reproduce the data quite well. The obtained parameters are summarized in Table I. For CeRu$_4$P$_{12}$, three Lorentzians were necessary: two Lorentzians for peak $D$ and one Lorentzian for peak $E$. Double Lorentzian in peak $D$ corresponds to excitation into $e_g$ and $t_{2g}$ band of $5d$ orbitals, which is split by crystal field mainly of twelve P atoms surrounding the Ce atom. On the other hand, to fit spectrum of PrRu$_4$P$_{12}$ two Lorentzians for peak $F$ were sufficient. It seems that there is no extra intensity around the satellite position of peak $F$. The splitting in peak $F$ is also the result of crystal-field effect.

To study temperature dependence of peak shape with high accuracy, it is better to reduce the number of variables as much as possible. We fitted peak $F$ of PrRu$_4$P$_{12}$ using one Lorentzian with linear background as

$$F(\omega) = \frac{a}{(\omega - \varepsilon_p)^2 + (\gamma/2)^2} + b\omega + c, \quad (2)$$

where $\varepsilon_p$ and $\gamma$ depict peak position and linewidth, respectively. Parameters $a$, $b$, and $c$ are constant values. The values of $b$ and $c$ are fixed at all temperatures. As is shown in the inset of Fig. 3, the fitted curve reproduces the data in the energy range 6439 eV $\leq \omega \leq$ 6455 eV quite well. Temperature dependence of the obtained Lorentzian linewidth is depicted in Fig. 3. The linewidth keeps constant and shows no anomaly even at $T = T_{MI}$. The result indicates that the shape of peak $F$ is independent of temperature.

## IV. DISCUSSION

As shown above, Ce $L_3$ edge of CeRu$_4$P$_{12}$ has a well-defined peak with a satellite. According to a previous report,[21] a similar satellite peak is observed in CeFe$_4$P$_{12}$ but not in LaFe$_4$P$_{12}$ and PrFe$_4$P$_{12}$. The satellite is also unobservable in the present experiments for PrRu$_4$P$_{12}$. That is, intensity of the satellites strongly depends on composition. We note that the peak position of the satellites may coincide with the beginning of extended x-ray absorption fine-structure (EXAFS) oscillations with the nearest-neighbor pairs, $L$-P ($L$ = rare earth), with a bond length of about 3.1 Å.[4,22] However, the satellites, observed only in a specific composition, cannot be an EXAFS signal. The EXAFS pattern should not be drastically different among those isostructural compounds with similar lattice constant. Therefore, we conjecture that these satellite peaks have electronic origin.

Taking account of the strong hybridization between Ce $4f$ and ligand electrons, ground state can be described as $|\psi\rangle = m|4f^1\underline{L}\rangle + n|4f^0\rangle$. In this case, when $2p$ core-state electrons are excited by x-ray photon to $5d$ states, final states split into $2p4f^1\underline{L}5d^*$ and $2p4f^05d^*$ states as observed in CeO$_2$. Magnitude of energy split in CeO$_2$ is about 8 eV, which is almost consistent with that between main and satellite peaks of CeRu$_4$P$_{12}$. This agreement supports the as-

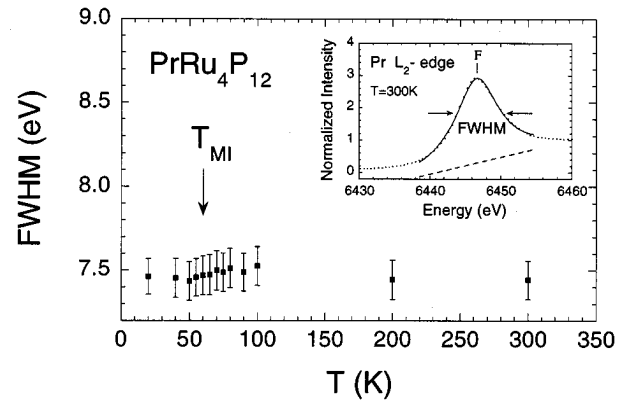

FIG. 3. Temperature dependence of resonance peak linewidth estimated by fitting using one Lorentzian function and a linear function in Pr $L_2$ edge for PrRu$_4$P$_{12}$. The fitting line is depicted by the solid line in the inset.

sumption that complex spectrum of $CeRu_4P_{12}$ also originates from strong hybridization of Ce $4f$ electrons in ground state. The main and satellite peaks of $CeRu_4P_{12}$ may correspond to $2p4f^1L5d^*$ and $2p4f^05d^*$ states, respectively. Extremely larger intensity of the main peak than the satellite suggests that Ce is mainly trivalent.

According to powder x-ray measurements, lattice constant of skutterudite compounds changes smoothly with atomic number of rare earth except for $CeRu_4P_{12}$ with an anomalously small value.[3,22] It was conjectured that valency of Ce is tetravalent different from other trivalent rare-earth skutterudite. Our present results, on the other hand, indicate that Ce is trivalent, and the small lattice constant in $CeRu_4P_{12}$ might be the result of strong hybridization.

Band calculations based on both the itinerant[6] and localized[5] $f$ electron models can explain semiconducting transport property in $CeRu_4P_{12}$ as described in the Introduction. In terms of Ce valency, however, only the former itinerant model is compatible with the observed trivalent Ce state. In the itinerant model, the semiconducting gap is formed by strong hybridization of Ce $4f$ with Ru $4d$ and P $3p$ electrons, which is also consistent with our observation of the satellite interpreted as a result of hybridization. Thus, we have concluded that the band gap of $CeRu_4P_{12}$ is a hybridization gap.

In Pr$L_2$ edge of $PrRu_4P_{12}$, a single peak without satellite is observed in the temperature range $20\,\text{K} \leq T \leq 300\,\text{K}$. The lack of satellite peak suggests that Pr is trivalent with completely localized $4f$ electrons, which play no role in conduction. The fact that Pr is trivalent in all temperature ranges leads to a conclusion that the MI transition at $T_{MI} = 60\,\text{K}$ does not originate from Pr valence fluctuation. Since XANES spectrum is sensitive to change in crystal field, the absence of temperature dependence in the spectrum also suggests that the microscopical environment around Pr atoms is stable at all temperatures. We could not find any role of Pr in the MI transition.

From a microscopical point of view, local lattice distortion around Ru and P atoms without affecting the nearest neighbors of Pr atoms remains as a possible origin of the MI transition. For example, Ru-P or P-P bond length distortions are another candidate for disturbing conduction of carriers. However, these distortions are undetectable in our Pr XANES measurements. Further microscopic study is required for clarifying mechanism of the MI transition.

## V. CONCLUSION

We have studied valency of rare earth in $LRu_4P_{12}$ ($L$ = Ce and Pr) by $L_{2,3}$-edge XANES spectroscopy. Spectrum of Ce $L_3$-edge shows a well-defined peak with a satellite that indicates that Ce is mainly trivalent with strongly hybridized $4f$ electrons. Band gap of $CeRu_4P_{12}$ seems to be formed by the strong hybridization. Pr $L_2$-edge spectra show a single peak structure without any change in the temperature range of $20\,\text{K} \leq T \leq 300\,\text{K}$, which indicates that valency of Pr is trivalent at all temperatures and the MI transition does not originate from Pr valence fluctuation.

## ACKNOWLEDGMENTS

The authors thank I. Hase and N. Shirakawa for delightful discussions and H. Unoki for sample preparation of $CeO_2$ and $CeTiO_3$. The synchrotron radiation experiments were performed at the SPring-8 (Proposal No. 1998A0293-NX-np) and at the Photon Factory. This work was supported by special promotion funds of the Science and Technology Agency, Japan.